\documentclass[aps,10pt,prl,tightenlines,twocolumn,twoside,showpacs,superscriptaddress,nofootinbib]{revtex4-1}

\usepackage{amsfonts}
\usepackage{float}
\usepackage{placeins}
\usepackage{graphicx}
\usepackage[hidelinks]{hyperref}
\usepackage{color,amsmath,amssymb,bm}
\usepackage{tensor}

\usepackage{lineno}

\usepackage{color}
\definecolor{dgreen}{cmyk}{1.,0.,1.,0.2}        
\definecolor{orange}{cmyk}{0.,0.353,1.,0.}    

\def \a {\alpha}
\def \b {\beta}

\def \e {\varepsilon}

\def \r {\rho}
\def \l {\lambda}
\def \m {\mu}
\def \n {\nu}
\def \s {\sigma}

\def \t {\tau}

\def \D {\Delta}

\def \L {\Lambda}
\def \vp {\bm{p}}

\def \CA {{\cal A}}

\def \SHE {$\nabla\mu_{B}$-IP}
\def \bSHE {baryonic SHE}
\newcommand\sect[1]{\noindent \textbf{#1}---}

\begin{document}

\title{Signatures of a baryonic spin-Hall effect in dense QCD matter}%
\author{Baochi Fu}
\email{fubaochi@pku.edu.cn}
\affiliation{Center for High Energy Physics, Peking University, Beijing 100871, China}
\affiliation{Department of Physics and State Key Laboratory of Nuclear Physics and Technology, Peking University, Beijing 100871, China}
\affiliation{Collaborative Innovation Center of Quantum Matter, Beijing 100871, China}
\author{Long-Gang Pang}
\email{lgpang@mail.ccnu.edu.cn}
\affiliation{Key Laboratory of Quark and Lepton Physics (MOE) and Institute of Particle Physics, Central China Normal University, Wuhan 430079, China}
\author{Huichao Song}
\email{huichaosong@pku.edu.cn}
\affiliation{Department of Physics and State Key Laboratory of Nuclear Physics and Technology, Peking University, Beijing 100871, China}
\affiliation{Collaborative Innovation Center of Quantum Matter, Beijing 100871, China}
\affiliation{Center for High Energy Physics, Peking University, Beijing 100871, China}
\author{Yi Yin}
\email{yiyin@cuhk.edu.cn}
\affiliation{
School of Science and Engineering, 
The Chinese University of Hong Kong (Shenzhen), 
Longgang, Shenzhen, Guangdong, 
518172, China
}

\date{\today}%

%
%
%
%
%
\begin{abstract}

Recent theoretical analyses show that spin current can be induced by the baryon chemical potential gradient $\nabla \mu_{B}$ which becomes sizable in the fireballs created in heavy-ion collisions (HIC) at a beam energy of ${\cal O}(10)$~GeV. 
This spin transport phenomenon can be considered as the cousin effect of the spin Hall effect (SHE) in condensed matter systems with $\nabla \mu_{B}$ playing the role of the analogous electric field.
In this letter,
we study this important mechanism, which we call ``Baryonic SHE" or $\nabla\mu_{B}$-induced polarization (\SHE),  for differential spin polarization generation that has not been systematically explored before.
We predict the signature of the baryonic SHE in HIC using a (3+1)~D viscous hydrodynamic model MUSIC with AMPT initial condition. 
We propose to use the second Fourier coefficients of the net spin polarization of Lambda hyperon as sensitive probes for the baryonic SHE.
Those baryonic SHE observables show a qualitative difference in sign and beam energy dependence for the situations with and without the baryonic SHE. 
Future experimental observation of these distinct qualitative features would provide strong evidence for the existence of this "Baryonic SHE" in the hot and dense QCD matter. 

\end{abstract}
\maketitle

%
\sect{Introduction}
The recent observation of spin polarization of hadrons in relativistic heavy-ion collisions, in particular that of $\Lambda$ hyperon (and $\overline{\Lambda}$ anti-hyperon)~\cite{STAR:2017ckg,Voloshin:2017kqp,STAR:2018gyt,ALICE:2019onw,STAR:2019erd,ALICE:2021pzu}, has inspired many studies on spin effects in QCD many-body systems (see Refs.~\cite{Voloshin:2017kqp,Becattini:2020ngo,Becattini2022} for reviews).
While the ``global" (phase space averaged) spin polarization is well-described by the effects of vorticity for a wide range of beam energy, 
the differential polarization measurements of the azimuthal angle dependence at top RHIC and LHC energies~\cite{STAR:2019erd,ALICE:2021pzu} show sign opposite to the theory based on thermal vorticity effects~\cite{Becattini:2017gcx}. 
Newly-discovered shear-induced polarization (SIP)~\cite{Liu:2021uhn,Fu:2021pok,Becattini:2021suc} is shown to be crucial for solving this ``polarization sign puzzle"~~\cite{Fu:2021pok,Becattini:2021iol}.
The examples of studying differential polarization at BES energies can be found in Refs.~\cite{Karpenko:2016jyx,Xie:2016fjj,Xia:2018tes,Wei:2018zfb,Fu:2020oxj,Guo:2021udq,Sun:2021nsg}.

Despite those developments,
much less attention has been paid to the quantitative signature of another important mechanism for spin polarization generation, namely, polarization induced by the gradient of reduced baryon chemical potential $\nabla \mu_{B}/T$, i.e., \SHE .
That is, for a spin carrier of momentum $\vp$ with baryon charge $q_{B}$, the induced spin polarization vector $\bm{P}$ is given by \footnote{Note that $\nabla\mu_B$-IP represents the contribution from the gradient of the reduced chemical potential $\mu_{B}/T$, and thus includes the term proportional to the temperature gradient multiplied by the baryon charge $q_B$.}
\begin{align}
\label{SHE}
    \bm{P}\propto - \vp\times (q_{B}{\bm \nabla}\,\mu_{B}/T)\, .
\end{align}
This phenomena has been predicted within the framework of relativistic quantum kinetic equations~\cite{Son:2012zy,Hidaka:2016yjf,Hidaka:2017auj,Hattori:2019ahi,Liu:2021uhn,Yi:2021ryh} and linear response theory~\cite{Liu:2020dxg,Liu:2021uhn}. 
Eq.~\eqref{SHE} shows that spin polarized by the \SHE~depends on momentum. 
Therefore one should look for the \SHE~signature in the phase space distribution of polarization observables, i.e. the ``local" spin polarization, rather than in the global polarization.

In this letter,
we present the first quantitative prediction of the \SHE~signals in differential spin polarization that can be tested experimentally at RHIC beam energy scan (BES) energies.
Using the freeze-out profiles generated from MUSIC viscous hydrodynamics with AMPT initial conditions,
we calculate the spin polarization of $\Lambda$ and $\overline{\Lambda}$ hyperons along the beam direction $\hat{z}$ and the out-plane direction $\hat{y}$ in Au+Au collisions with the beam energy   $\sqrt{s_{NN}} = 7.7-200$~GeV.
Unlike the previous work~\cite{Liu:2020dxg} which only accounted for \SHE~alone within the blastwave model, 
we have included all first-order gradients effects in our studies. 
As we shall see later, such a comprehensive analysis is necessary to identify the signature of \SHE.
In recent hydrodynamic calculations, the \SHE~contribution on global polarization is studied and the magnitude is found to be relatively small compared with the well-known thermal vorticity effect~\cite{Ryu:2021lnx}. While in this work, we demonstrate that the \SHE~gives rise to significant separation for the differential polarization between $\Lambda$ and $\overline{\Lambda}$.
The second harmonics of the net spin polarization can be utilized as unambiguous observables to search for the \SHE.

It is worth pointing out that if we replace $\nabla\mu_{B}/T$ with the electric field in Eq.~\eqref{SHE},
the resulting equation describes the spin Hall effect (SHE)~\cite{DYAKONOV1971459,sinova2014spin}. 
In this sense, 
\SHE~is an analogue of SHE which has been observed in semiconductors~\cite{doi:10.1126/science.1105514,PhysRevLett.94.047204}, metals~\cite{PhysRevLett.106.036601} and insulators~\cite{2007Science}; see Ref.~\cite{sinova2014spin} for a review.
However, the materials mentioned above are at or below room temperature and are microscopically described by QED.
In contrast, we consider the signature of \SHE~in hot and dense QCD matter where
the spin carriers (quarks, gluons and/or hadrons) interact with each other through strong interaction in media at temperature above $10^{12}$~K. 
Those similarity and difference between SHE and \SHE~makes the detection of the latter of broader interest.

\sect{Method}
In field theory, the phase space density of spin polarization of a fermion/anti-fermion with mass $m$ and four-momentum $p^{\mu}$ is described by the corresponding axial Wigner function $\CA^{\mu}(x,p)$.
To the first order in the gradient of hydrodynamic variables, i.e., temperature $T$, flow velocity $u^\m$, chemical potential $\m_B$,
$\CA^{\mu}$ for a baryon-rich fluid is given by~\cite{Liu:2021uhn,Liu:2020dxg} (see also Refs.~\cite{Hidaka:2017auj,Yi:2021ryh}):
\begin{widetext}
\begin{equation}
\label{eq:spin}
{\CA}^\mu (x, p) = \b f_0(x,p) (1 - f_0(x,p))  \e^{\m\n\a\r} \times \Big(
                   \underbrace{\frac{1}{2_{}} p_\n \partial_{\a}^\perp u_\r}_{\text{vorticity}}
                 - \underbrace{\frac{1}{T} u_\n p_{\a} \partial_\r T }_{\text{T-gradient}}
                 - \underbrace{\frac{p^2_\perp}{\e_0}  u_\n Q^\l_\a \s_{\r\l}}_{\text{SIP}}
                 - \underbrace{\frac{q_{B}}{\e_0 \b} u_\n p_{\a} \partial_{\r}(\b \m_B)}_{\text{Baryonic SHE}}
                   \Big),
\end{equation}
\end{widetext}
where $ f_0(x,p) = (e^{(\e_0 - q_{B} \m_B)\beta} + 1)^{-1}$ is the Fermi-Dirac distribution function with $\e_0 = p \cdot u$ and $\beta=1/T$.
Note $q_{B}=1$ for baryons and $q_{B}=-1$ for anti-baryons.
Here, we define $\partial^\m_\perp \equiv \D^{\m\n} \partial_\n, p_{\perp}^{\mu}\equiv \D^{\mu\nu}p_{\nu}$ with $\D^{\m\n} = g^{\m\n} - u^\m u^\n$.
We denote momentum quadrupole by $Q^{\m\n} = -p^\m_\perp p^\n_\perp / p^2_\perp + \D^{\m\n} /3$ and shear stress tensor by $\s^{\m\n} = \partial^{(\m}_\perp u^{\n)}_{{}} - (1/3)\D^{\m\n} (\partial\cdot u)$.

\begin{figure*}[t]
\includegraphics[
trim={2cm 0cm 2cm 0cm},clip,
width=0.99\textwidth]{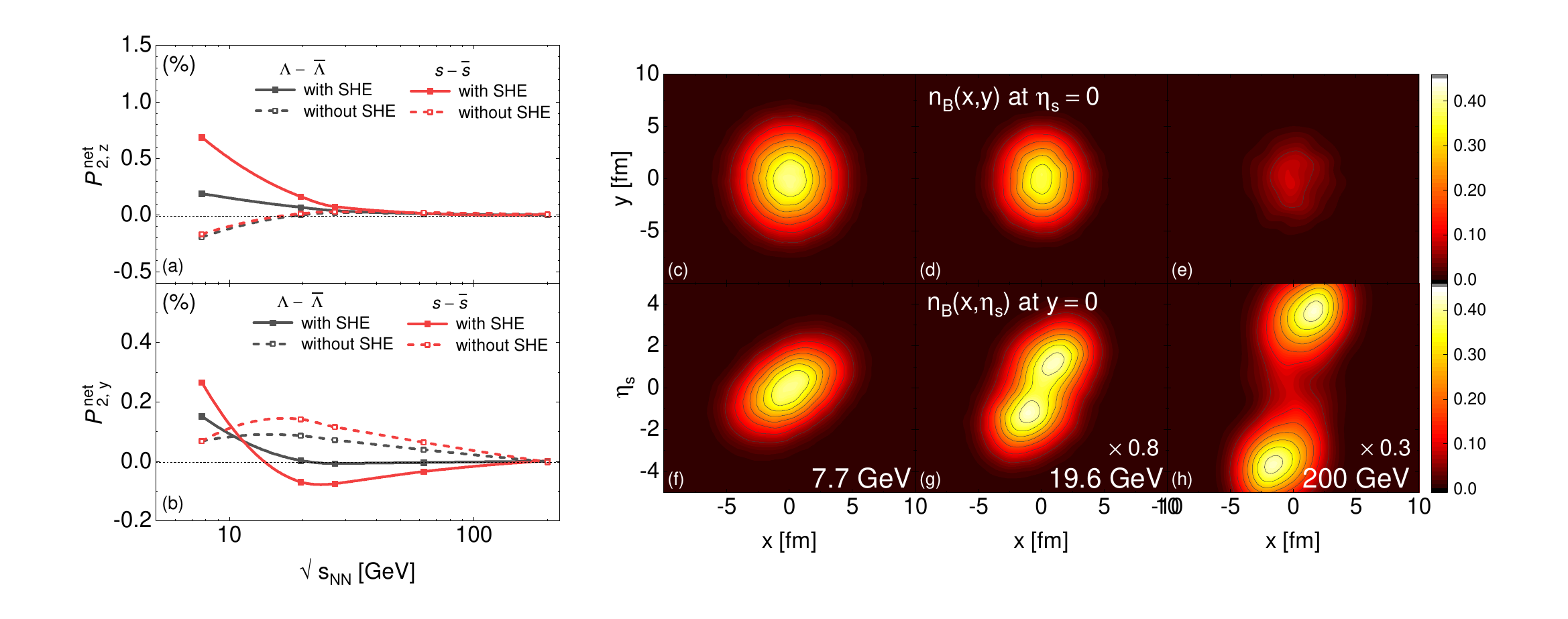}
%
%
\caption{
\label{fig:main-result} Left panels: the collision energy dependence of second Fourier coefficients, $P^{{\rm net}}_{2,z}$ and $P^{{\rm net}}_{2,y}$, for differential polarization of net $\Lambda$ hyperons along the beam and out of plane direction, calculated in the ``Lambda equilibrium" and ``strange memory" scenarios, see text.
Right panels: the initial net baryon density  $n_{B}$ on the transverse plane $x-y$ at spatial rapidity $\eta_s=0$ and on the reaction plane $x-\eta_{s}$ at $y=0$ in 20-50\% Au+Au collisions at $\sqrt{s_{NN}} = 7.7,19.6,200$~GeV. 
}
\end{figure*}

Eq.~\eqref{eq:spin} includes all possible first order hydrodynamic gradient contributions to the spin polarization in phase space that are allowed by symmetry,
except for the vorticity-quadrupole coupling which is found to be suppressed in perturbative calculations~\cite{Liu:2021uhn}.
Among those four independent contributions, 
the first and second terms in the brackets represent the vorticity-induced and T-gradient induced polarization respectively.
The combination of the two is proportional to the thermal vorticity, the effect of which has been widely studied previously~\cite{Becattini:2013fla,Becattini:2017gcx}.
The recently-discovered shear-induced polarization (SIP)~\cite{Liu:2021uhn,Fu:2021pok,Becattini:2021suc,Yi:2021ryh} is shown in the third term. 
The \bSHE~contribution, given by the last term, is of potential importance for baryon-rich QCD matter.  
A similar expression for the baryonic SHE has been derived using a statistical approach \cite{Buzzegoli:2022kyj}. This approach differs from Eq.~(\ref{eq:spin}) primarily due to the inclusion of the four-velocity vector \( u^\mu \) and a time-direction vector \( \hat{t}^\mu \). In a more recent study \cite{Sheng:2025cjk}, these vectors have been replaced by the unit surface normal vector \( n^\mu \).
Equation (\ref{eq:spin}) was derived using a free theory analysis and includes four distinctive gradient contributions. We anticipate that higher-loop contributions may modify the coefficients in front of each gradient term, but they will not change the presence or qualitative features of these contributions.
We shall employ Eq.~\eqref{eq:spin} as a basis for the subsequently systematic analysis of spin polarization at BES energies
\footnote{Note the expression of Eq.~(\ref{eq:spin}) is not unique, which can be equivalently written as the combination of derivatives of flow velocities and temperature fields in ideal hydrodynamics. With the convention of \cite{Baier:2007ix}, we eliminate the time derivatives and focus on the effects induced by the $\mu_B$ in this work. Nevertheless, the value of the predicted SHE signature does not depend on a specific form.}.

To compute the spin polarization of the produced $\L$ and $\overline{\Lambda}$ hyperons on the freezeout-surface $\Sigma_{\mu}$ from the axial Wigner function,
we follow the widely-used prescription~\cite{Becattini:2013fla, Fang:2016vpj}:
\begin{equation}
    P^\m(\bm{p}) = \frac{\int{d \Sigma^\a p_\a \CA^\m(x,\bm{p};m) }} {2m \int{d \Sigma^\a p_\a f_0(x,p)}}\, ,
\label{eq:frz}
\end{equation}
where the factor $2$ in the denominator comes from the degeneracy of spin $1/2$ fermions.

\begin{table}
\begin{tabular}{ |c||c|c|c|c| }
 \hline
$\sqrt{s_{NN}} [{\rm GeV}]$  & $c_{N}$ & $\eta/s$  & $\t_0 {\rm [fm/c]}$ & $\epsilon_\text{SW}\ {\rm [GeV/fm^3]}$ \\
 \hline
 7.7  & 1.8 & 0.16 &  2.0 &   0.35\\
 \hline
 19.6 & 1.8 & 0.12 &  1.2 &  0.45\\
 \hline
 27    & 1.8 & 0.10 & 1.0  &  0.50\\
 \hline
 62.4  & 1.9 & 0.08 & 0.6  &  0.60\\
 \hline
200   & 1.9 & 0.08 &  0.4 &  0.65\\
 \hline
\end{tabular}
\caption{The subset of parameters used in hydrodynamic simulations,
see text.}
 \label{tab:para}
\end{table}

The hydrodynamic profile on the freeze-out surface is generated by 3+1 dimensional viscous hydrodynamics MUSIC. 
The initial energy-momentum tensor $T^{\mu\nu}(x)$ and net baryon number density $n_B(x)$ are obtained from AMPT model~\cite{Lin:2004en, He:2017tla} with Gaussian smearing at proper time $\tau_0$~\cite{Pang:2012he} after average over more than $10^{5}$ events at each $\sqrt{s_{NN}}$. 
We have multiplied $T^{\mu\nu}$ obtained from AMPT by a phenomenological normalization factor $c_{N}$ which is tuned to match with charged hadron production in most central collisions. 
The hydrodynamic evolution ends when energy density reaches a fixed value $\e_{\rm SW}$. 
Table~\ref{tab:para} lists relevant parameters used in our calculation of hydrodynamic profiles.
Unless noted in Table.~\ref{tab:para}, the model set-up for AMPT+MUSIC are the same as our previous studies~\cite{Fu:2020oxj, Fu:2021pok}.

Given significant uncertainties in the current understanding of the hadronization of spin polarization and its subsequent hadronic evolution, we consider two benchmark scenarios: the ``Lambda equilibrium" and ``strange memory"~\cite{Fu:2021pok}.
The former assumes that the spin of $\L$ immediately responds to the hydrodynamic gradients~\cite{Pang:2016igs, Karpenko:2016jyx, Fu:2020oxj}. 
In contrast, the latter supposes that the spin of $\L$ inherits that of the strange quark and is frozen after the hadronization~\cite{Liang:2004ph, Gao:2020lxh,Liu:2019krs}.
In practice, we input $m= 1.116 \;\textrm{GeV}, q_{B}=\pm 1$ and $m = 0.3 \;\textrm{GeV},q_{B}=\pm 1/3$ in Eq.~\eqref{eq:frz} for the ``Lambda equilibrium" and ``strange memory" scenario respectively, as we did in Ref.~\cite{Fu:2021pok}.
Clearly, those two scenarios represent two extremes, i.e., the spin relaxation time of $\Lambda$ is very short $\tau_{{\rm spin}}\rightarrow 0$ and very long $\tau_{{\rm spin}}\rightarrow \infty$, respectively, with the reality (possibly) sitting somewhere in between.
Note, Eq.~\eqref{eq:spin} is derived by assuming that spin carriers are weakly interacting with each other~\cite{Liu:2020dxg,Liu:2021uhn} while for QGP near the freezeout temperature, the strange quark may be strongly coupled the medium.  
In particular for RHIC-BES collisions, with the decreasing of collision energy, we may expect the spin of $\Lambda$~hyperons is more likely to exhibit the hadron spin degree of freedom instead of quarks.
Nevertheless, as we shall see below, the main feature of the SHE observables that we propose is robust in both scenarios.


\sect{Results}
To devise the \bSHE~observables, we shall rely on two observations followed from Eq.~\eqref{eq:spin}.
First, 
as evident in this equation, 
the \bSHE~distinguishes itself from the other contributions there since its sign depends on the baryon charge of the spin carrier.
It is then natural to consider net $\Lambda$ differential spin polarization $P^{{\rm net}}_{z,y}(\phi) \equiv P_{z,y}(\phi)-\overline{P}_{z,y}(\phi)$ to isolate the \bSHE~contribution. 
Here $P_{z,y}(\phi)$ and $\overline{P}_{z,y}(\phi)$ is the polarization of $\L$ and $\overline{\L}$ hyperons as a function of azimuthal angle $\phi$ along the out-of-plane direction ($y$) and the beam direction ($z$).
Second, we notice that the \bSHE~contribution is of a similar form to that of the T-gradient contribution with $\partial_{\mu} T$ replaced by $(q_{B} T^2/\e_{0})\partial_{\mu} (\beta\mu_{B})$. 
It has been demonstrated in a number of studies~\cite{Becattini:2017gcx,Fu:2020oxj,Fu:2021pok} that the effect of T-gradient will lead to a characteristic ``$\sin(2\phi)$" and ``$\cos(2\phi)$" pattern to $P_{z}(\phi)$ and $P_{y}(\phi)$ respectively. 
So we expect that the \bSHE~would give rise to a similar sine/cosine pattern in differential polarization, with the sign depending on the typical direction of $\mu_{B}$ gradient on the freezeout surface.  
Combining those observations, we designate the second harmonic component of $P^{{\rm net}}_{z,y}(\phi)$ in the Fourier decomposition to characterize the \bSHE~signal:
\begin{align}
\label{Obervables}
    P^{{\rm net}}_{2,z}\equiv \langle P^{{\rm net}}_{z}(\phi)\sin2\phi\rangle\, ,
    P^{{\rm net}}_{2,y}\equiv -\langle P^{{\rm net}}_{y}(\phi)\cos2\phi\rangle\,
\end{align}
where $\langle...\rangle$ denotes the average over the $\phi$ angle.

\begin{figure*}[t] 
\includegraphics[
trim={2cm 0cm 2cm 0cm},clip,
width=1.95\columnwidth]{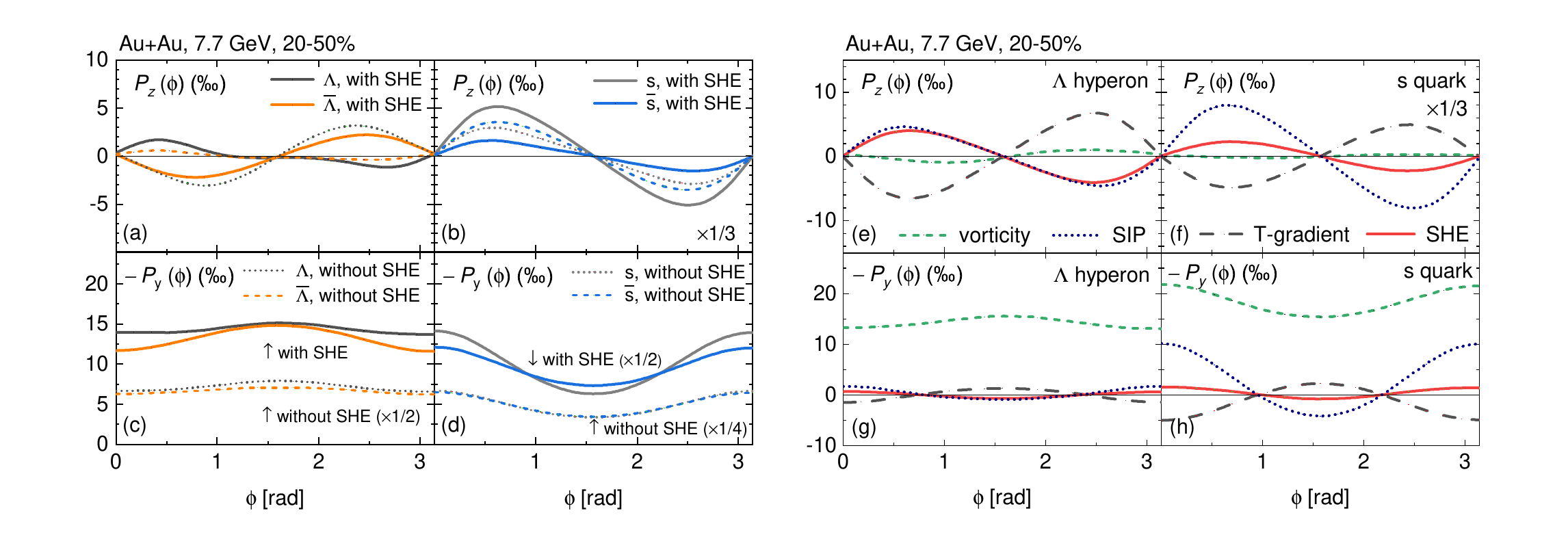}
\caption{
\label{fig:local}  
Differential spin polarization $P_z(\phi)$ and $-P_y(\phi)$ calculated in the ``Lambda equilibrium" (labeled by $\Lambda,\overline{\Lambda}$) and the ``strange memory" (labeled by $s,\bar{s}$) scenarios in 20-50\% Au+Au collision at $\sqrt{s_{NN}} = $ 7.7 GeV.
Left panels:
with and without the SHE contribution for $\L$ and $\overline{\L}$ hyperons.
Right panels: The individual contribution from the vorticity, the T-gradient, the shear-induced polarization (SIP) and the SHE terms in Eq.~\eqref{eq:spin} to the polarization of $\L$.
}
\end{figure*}

The main results of this letter are shown in Fig.~\ref{fig:main-result} (left) where the proposed \bSHE~observables $P^{{\rm net}}_{2,z}$ and $P^{{\rm net}}_{2,y}$ as a function of collision  energy are plotted for both ``Lambda equilibrium" and ``strange memory" scenarios.
While the magnitude of those observables quantitatively depends on the mass of spin carriers, 
we observe several distinctive traits which are robust in both scenarios.
Without the \bSHE, the second sine harmonics of longitudinal net spin polarization $P^{{\rm net}}_{2,z}$ is negative at the beam energy of ${\cal O}(10)$~GeV. 
With the \bSHE, it becomes positive and grows as collision  energy decreases. 
Turning to the \bSHE~observable in $y$-direction, we see a different non-monotonic collision  energy dependence in situations with and without the \bSHE. 
As beam energy goes down, $P^{{\rm net}}_{2,y}$ shows an interesting sign change, from negative to positive, in the presence of the \bSHE. 
In contrast, this cosine coefficient would first increase to some positive value and then drop down when the \bSHE~is absent. 
In short, we find that the proposed \bSHE~observables show a marked difference in both sign and beam energy dependence with and without the \bSHE. 
Those striking qualitative features make us believe the predicted signatures provide crucial guidance for the future search for the \bSHE.

Physically, the \bSHE~converts the spatial baryon chemical potential distribution into spin polarization in momentum space. 
Therefore, the characteristics of the baryon density profiles and their beam energy dependence would appear in observables sensitive to the \bSHE. 
To illustrate this,  in Fig.~\ref{fig:main-result} (right), we present initial baryon density distribution, 
which evolves hydrodynamically and can be converted to $\mu_{B}$ through EoS. 
As Eq.~\eqref{SHE} tells us, the spin vector polarized by the \bSHE~is transverse to the gradient of $\mu_{B}$.
So, to appreciate the behavior of $P^{{\rm net}}_{2,z}$, 
we need to look at the $n_{B}$ distribution in the $x-y$ plane. 
We notice that the magnitude of $n_{B}$ becomes larger with decreasing beam energy. 
Accordingly, we see the growing trends of $P^{{\rm net}}_{2,z}$ when beam energy becomes smaller. 
Turning to the initial profile of $n_{B}$ in the reaction plane $x-\eta_{s}$ at $y=0$,
it transits from a double peak structure to a single peak structure from $\sqrt{s_{NN}}=200$~GeV to $7.7$ ~GeV. 
This transition is anticipated from examining baryon stopping process at different collision energies~\cite{Lin:2004en,Videbaek:2009zy, NA49:2010lhg, Busza:2018rrf, Mohs:2019iee}. 
Such a qualitative change in baryon density profile will flip the sign of baryon density gradient along the longitudinal direction and is expected to induce the non-monotonic behaviors of $P^{{\rm net}}_{2,y}$, as was confirmed quantitatively in Fig.~\ref{fig:main-result} (left) 
\footnote{
When computing SHE quantitatively, 
we need to evaluate $\nabla(\mu_{B}/T)$ on the freezeout surface which is related to $n_{B}$ through $T$ profile and EoS. 
Therefore, Fig.~\eqref{fig:main-result} (right) is presented only for illustrative purpose. 
Moreover, we notice that although $\nabla n_{B}$ can be expressed in terms of a linear combination of $\nabla T$ and $\nabla \mu_{B}$, 
the sign of $\nabla n_{B}$ is not always the same as that of $\nabla \mu_{B}$ because of non-trivial $\nabla T$ profile, see Ref.~\cite{Fu-future} for details.
}.

It is worth emphasizing that even without the \bSHE, $P^{{\rm net}}_{2,z}, P^{{\rm net}}_{z,y}$ is generically non-zero at BES energies. 
This is because there are other mechanisms for the spin polarization, such as the SIP, with magnitude depending on $\mu_{B}$ through distribution function and Pauli blocking factor $(1-f)$ in Eq.~\eqref{eq:spin}. 
An important factor within the present model is that $\L$ tends to freeze-out earlier than $\overline{\L}$ , meaning $\L$ and $\overline{\L}$ probe the regimes with different hydrodynamic profiles. 
The resulting non-trivial non-SHE contribution necessitates a quantitative description of differential spin polarization by systematically including both the \bSHE~and non-SHE effects, as was done for the first time in the present letter.

To complement Fig.~\ref{fig:main-result}, 
we show $P_z(\phi), P_y(\phi)$ and $\overline{P}_z(\phi),  \overline{P}_y(\phi)$ at one representative beam energy $\sqrt{s_{NN}}=7.7$~GeV for both ``$\Lambda$ equilibrium'' and ``strange memory'' scenarios in Fig.~\ref{fig:local} (left). 
This figure exemplifies several characteristic features which are also present at other beam energies, see our coming publication for details~\cite{Fu-future}. 
First, 
azimuthal angle dependence of $P_{z,y}(\phi)$ ($\overline{P}_{z,y}(\phi)$) is mainly characterized by $\cos(2\phi), \sin(2\phi)$ pattern.
This supports choosing the second harmonics of differential polarization to detect the \bSHE. 
Second, the \bSHE~leads to a sizable separation of $\Lambda$ and $\overline{\Lambda}$ differential polarization. 
Without the \bSHE, either the difference between them is small, or the ordering is opposite to that with the \bSHE. 
For example, in ``$\Lambda$ equilibrium'' scenario, the splitting between $P_{y}(\phi)$ and $\overline{P}_{y}(\phi)$, shown in the bottom-left panel of Fig.~\ref{fig:local} (left), is small in the absence of the \bSHE~but becomes much larger once the \bSHE~contribution is included. 
Meanwhile, the hierarchy between $P_{z}(\phi),\overline{P}_{z}(\phi)$ will be reversed with and without the \bSHE. 
Similar qualitative features can equally be seen in the ``strange memory'' scenario, per the lessons we have just discussed when presenting Fig.~\ref{fig:main-result}.

In Fig.~\ref{fig:local} (right), 
we present each of the four contributions, i.e., from the SIP, the T-gradient and the vorticity in additional to the \bSHE, to $P_{y,z}$ at $\sqrt{s_{NN}}=7.7~$GeV. 
For longitudinal polarization $P_{z}(\phi)$, 
the vorticity contribution is insignificant. 
Similar to results of hydrodynamic calculations at top RHIC and LHC energies~\cite{Fu:2021pok,Becattini:2021iol}, 
the sign of the second sine harmonics of the SIP and the T-gradient contribution is opposite to each other at the lower beam energies. 
The sign of the \bSHE~contribution, as shown in red curve, is positive and is the same as that from the SIP. 
Overall, 
the sign of second sine harmonics of $P_{z}$ will be determined by the competition among the \bSHE, the SIP and the T-gradient induced polarization. 
Turning to the polarization along $y$-direction, 
the ``global'' polarization is non-zero mainly because of the vorticity effect. 
Once again, the sign of the second cosine harmonics is driven by the competition among contributions from different hydrodynamic gradients. 
We note that while the sign of the SIP, T-gradient and vorticity contribution at $\sqrt{s_{NN}}=7.7~$GeV is same as that at top RHIC and LHC~\cite{Fu:2021pok,Becattini:2021iol},
the sign of the \bSHE~contribution to $P_{y}(\phi)$ is sensitive to the $\mu_{B}$ profile in the longitudinal direction and hence can change from negative to positive with decreasing beam energy, as we discussed earlier. 
The main insight provided by Fig.~\ref{fig:local} (right) is that the \bSHE~contribution is indispensable for studying differential spin polarization at lower beam energies. 
The net spin polarization is more sensitive to the unique feature of the \bSHE~than $P_{z,y}$ or $\overline{P}_{z,y}$ because the former largely isolates the effects of chemical potential gradient from other mechanisms for polarization generation.

Besides results presented above, we have investigated the consequences of changing the inputs in our hydrodynamic models. 
They include the initial baryon density and flow profile, shear viscosity, bulk viscosity, baryon diffusion constant, EoS, and freeze-out condition.
We find that the \bSHE~signal is mostly sensitive to the initial profile but those variations within realistic range don't qualitatively change the conclusion in this letter; see the forthcoming publication for more details~\cite{Fu-future}. 

It is important to note that other sources may and do contribute to the net Lambda polarization: electro-magnetic field \cite{Guo:2019mgh, Guo:2019joy}, helical vortical effect \cite{Ambrus:2020oiw}, freeze-out condition \cite{Vitiuk:2019rfv} - to name just a few. We note that such effects are expected to decay shortly or only insignificantly contribute to longitudinal polarization. Though, the quantitative analysis of these contributions is a task well beyond this paper, but such an analysis should be pursued soon.


\sect{Summary}
In this letter, we predict the baryonic spin Hall effect (SHE) signature for the QCD matter created in relativistic heavy-ion collisions at RHIC-BES energies. Such \bSHE~arises from the gradient of baryon chemical potential and is an essential mechanism generating spin polarization in baryon-rich QCD that has not been fully explored before. 
Using the freeze-out profiles from MUSIC hydrodynamics with AMPT initial conditions, we demonstrate that the \bSHE~contribution to the differential Lambda polarization is comparable to those from the thermal vorticity and the shear-induced polarization (SIP) at lower collision energies.  
We propose to use the second harmonics of net spin polarization, $P_{2,z}^{\rm net}$ and $ P_{2,y}^{\rm net}$ (defined in Eq.~\eqref{Obervables}), to detect the \bSHE.
We predict that these proposed \bSHE~observables will show the qualitative difference in both sign and collision energy dependence in the absence and presence of the \bSHE. 
Future experimental observation of the anticipated non-trivial signatures could provide the first evidence for the \bSHE~in hot and dense QCD matter.\\



\acknowledgments
We gratefully acknowledge the contributions of Shuai Liu, who collaborated with us on this research project during its early stages.
We thank helpful discussions with Iurii Karpenko, Chun Shen and especially thank Xiangyu Wu for the code verification on the part of the results presented in this work.
This work was supported in part by the NSFC under grants No.~12575138 and No.~12247107 (H.S.),
No.~12147173 (B.F.) 
and No.~12435009 (L.P.).
YY acknowledges the support from NSFC under grant No.12175282 and by CUHK-Shenzhen University Development Fund under the Grant No. UDF01003791.
We also acknowledge the extensive computing resources provided by the Supercomputing Center of Chinese Academy of Science (SCCAS), Tianhe-1A from the National Supercomputing Center in Tianjin, China and the High-performance Computing Platform of Peking University.\\

\sect{Data Availability}
The data that support the findings of this article are openly available \cite{data-arxiv}. 

\bibliography{SHE_refs}
\end{document}